\documentclass[twocolumn,floatfix, nofootinbib,prd,preprintnumbers,showpacs,showkeys,superscriptaddress,longbibliography]{revtex4-2}

\usepackage{amssymb}
\usepackage[mathscr]{euscript}
\usepackage{amsmath}
\usepackage{graphicx}
\usepackage{dcolumn}
\usepackage{bm}
\usepackage{epsfig}
\usepackage{latexsym,mathrsfs}
\usepackage{graphicx}
\usepackage{color}
\usepackage{hyperref}
\usepackage{float}
\usepackage[thinlines]{easytable}
\usepackage{multirow}
\usepackage{diagbox}
\usepackage{varwidth}

\newcolumntype{M}[1]{>{\centering\arraybackslash}m{#1}}
\newcolumntype{N}{@{}m{0pt}@{}}

\newcommand*\diff{\mathop{}\!\mathrm{d}}

\hypersetup{
    colorlinks=true,
    linkcolor=red,
    citecolor=blue,
}

\usepackage{physics}
\usepackage{csquotes}
\usepackage{mathtools}
\usepackage[caption=false]{subfig}
\usepackage{hyperref}
\usepackage{url}
\usepackage{xcolor}
\usepackage{color}
\definecolor{amaranth}{rgb}{0.9, 0.17, 0.31}
\definecolor{purple(munsell)}{rgb}{0.62, 0.0, 0.77}
\definecolor{americanrose}{rgb}{1.0, 0.01, 0.24}
\definecolor{palatinateblue}{rgb}{0.15, 0.23, 0.89}
\definecolor{royalblue(web)}{rgb}{0.25, 0.41, 0.88}
\definecolor{hanpurple}{rgb}{0.32, 0.09, 0.98}
\definecolor{beaublue}{rgb}{0.74, 0.83, 0.9}
\definecolor{carminered}{rgb}{1.0, 0.0, 0.22}
\definecolor{brightpink}{rgb}{1.0, 0.0, 0.5}
\definecolor{vividviolet}{rgb}{0.62, 0.0, 1.0}

\definecolor{electron}{rgb}{1.0, 0.67, 0.22}
\hypersetup{ linktoc=all,
    colorlinks, linkcolor={palatinateblue},
    citecolor={brightpink}, urlcolor={amaranth}}

\newcommand{\hlch}[1]{#1}

\begin{document}

\preprint{FTPI-MINN-23-21}

\title{
%
    \hlch{Moving mirrors and event horizons in non-flat background geometry}
	}

\author{Evgenii Ievlev}
\email{ievlev9292@gmail.com}
\altaffiliation[On leave of absence from: ]{National Research Center “Kurchatov Institute”, Petersburg Nuclear Physics Institute, St.\;Petersburg 188300, Russia}
\affiliation{William I. Fine Theoretical Physics Institute, School of Physics and Astronomy,\\
University of Minnesota, Minneapolis, MN 55455, USA}
\affiliation{Physics Department \& Energetic Cosmos Laboratory, Nazarbayev University,\\
Astana 010000, Kazakhstan}

\begin{abstract} 

Moving mirrors have been used for a long time as simple models for studying various properties of black hole radiation, such as the thermal spectrum and entanglement entropy.
These models are typically constructed to mimic the collapse of a spherically symmetric distribution of matter \hlch{in the 
Minkowski background.}
We generalize this correspondence to the case of non-trivial background geometry 
\hlch{
and consider two examples, the Schwarzschild -- de Sitter black hole and the Ba\~nados--Teitelboim--Zanelli (BTZ) black hole.
In the BTZ case we were also able to show that this approach works for the spinning black hole which has only axial symmetry.
}



\end{abstract}

\maketitle

\section{Introduction}

It has been known for a long time that the vacuum in a quantum system can change its structure when the system undergoes a transition, either smooth or discontinuous. 
This implies that even if the system has started as a state with lowest possible energy and no particles, at late times particles may appear.

In the Heisenberg picture of quantum mechanics this is seen from the evolution of the Hamiltonian and its spectrum.
\hlch{Such an evolution means that, although the wave function $\ket{\Psi}$ is not changing, its decomposition into the energy eigenstates might not be the same at early and late times.
Recall that any Hilbert space vector $\ket{\Psi}$ can be decomposed in a sum over the spectrum of a Hamiltonian with certain coefficients. If the spectrum of the Hamiltonian changes, these coefficients are also modified, like coordinates under a change of the basis.}

\hlch{The upshot of this is that even if the initial state $\ket{\Psi}$ was the lowest energy eigenstate of} the early-times Hamiltonian, at late times it may have contributions from excited eigenstates.

In quantum field theory (QFT), one can analyze vacuum fluctuation of the quantum field.
Under the influence of, for example, a black hole horizon in 3+1d \cite{Hawking:1974sw} or a dynamical boundary condition in 1+1d \cite{DeWitt:1975ys,Davies:1976hi,Davies:1977yv}, these fluctuations can be amplified, which leads to a detectable energy/particle flux being produced.

The moving mirror model, or dynamical Casimir effect (DCE), in the simplest case is a QFT of a massless scalar \hlch{field 
with} a Dirichlet boundary condition (b.c.): the field is set to zero at the point where the mirror is located. 
The non-trivial dynamics occurs if the mirror is accelerated 
\cite{moore1970quantum,DeWitt:1975ys,Davies:1976hi,Davies:1977yv,walker1985particle,Ford:1982ct,carlitz1987reflections,Chen:2015bcg,AnaBHEL:2022sri,Good:2021asq,Chen:2020sir},
see also e.g. the textbooks \cite{Birrell:1982ix,Fabbri}.

It was quickly realized that the phenomena of particle production near a black hole horizon and in moving mirror model bear striking resemblance, and that this resemblance can be used for understanding certain aspects of the QFT in 3+1-dimensional curved spacetime by studying a simpler 1+1 flat spacetime model \cite{DeWitt:1975ys,Davies:1976hi,Davies:1977yv}.
Later a specific recipe was constructed \cite{Wilczek:1993jn} that relates a collapsing \hlch{lightlike shell} to a specific moving mirror trajectory.
However, this recipe is valid as-is only in asymptotically flat spacetimes with spherical symmetry \hlch{and with a flat region inside the shell.}
Two such examples have been considered in the literature, the Schwarzschild \cite{Good:2016oey} and Reissner–Nordstr\"om \cite{good2020particle} black hole mirror analogies.

There have also been attempts to apply this recipe to spacetimes that are not asymptotically flat \cite{Good:2020byh,Fernandez-Silvestre:2021ghq,Fernandez-Silvestre:2022gqn} 
or have only axial symmetry \cite{Good:2020fjz,Foo:2020bmv}.
However, as these examples violate the initial prescription assumptions (asymptotic flatness and spherical symmetry), there is no guarantee that the results will be sensible; and indeed, inconsistencies may arise \cite{Good:2018zmx}.

In this work we generalize the black hole -- mirror correspondence to the case when the spacetime is not asymptotically flat, \hlch{and the shell's interior region is not flat eigher}.
More specifically, we consider a black hole that is formed from a collapsing shell of null dust immersed in a spacetime that had non-vanishing curvature to begin with.
The thin shell approximation allows to capture the essential features of the process while keeping full tractability.
We were also able to demonstrate consistency of relaxing the spherical symmetry assumption: instead of requiring that the spacetime itself should be spherically symmetric, 
we only require that the spherical symmetry is restored in the $s$-wave sector of massless particles' geodesics.
This might explain the consistency of the results on the Kerr(-Newman) black holes \cite{Good:2020fjz,Foo:2020bmv,Rothman:2000mm}.

The paper is organized as follows.
\hlch{In Sec.~\ref{sec:setup} we review basic results on particle and energy production from moving mirrors and black holes.} Readers familiar with the subject can safely skip this Section.
In Sec.~\ref{sec:wilczek} we discuss the details of the black hole -- mirror correspondence in a flat background, while in Sec.~\ref{sec:curvedback} we generalize it to a curved background geometry.
Sec.~\ref{sec:examples} presents two examples, the Schwarzschild -- de Sitter and the BTZ black holes ($s$-wave angular symmetry is discussed in subsection~\ref{sec:spin}).
In Sec.~\ref{sec:conclusions} we outline the main results and possible future directions.

\section{ Review of  particle production in dynamical models }
\label{sec:setup}

\hlch{
In this section we will briefly review the concept of particle production in some models with externally varied conditions.
This Section focuses on the key features of moving mirrors and black holes that would help us to understand the connection between moving mirrors and collapsing geometries.
}

\subsection{Moving mirror in 1+1d}
\label{sec:setup_mirror}

\begin{figure*}[t]
    \subfloat[Timelike mirror has no horizon.]{\label{fig:mirror_timelike}\includegraphics[width=0.8\columnwidth]{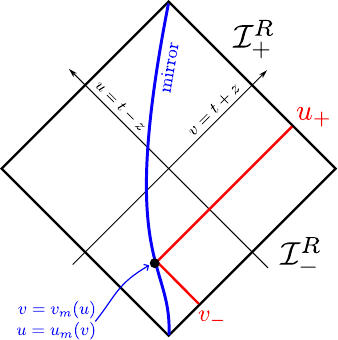}}
    \qquad
    \subfloat[Asymptotically lightlike mirror develops a horizon $\mathcal{H}$.]{\label{fig:mirror_lightlike}\includegraphics[width=0.8\columnwidth]{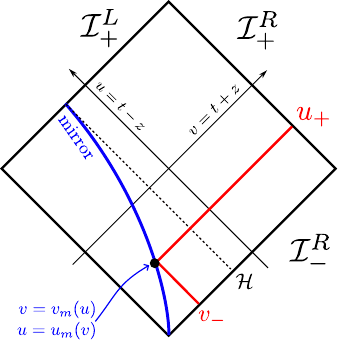}}
\caption{
    \hlch{
	Two Penrose diagrams for 1+1 mirrors with and without horizons.
	An incoming light ray starts at the past lightlike infinity $\mathcal{I}^R_-$ and propagates at $v = v_- = \text{const}$, then gets reflected at the mirror and propagates to the future lightlike infinity $\mathcal{I}^R_+$ with $u = u_+ = \text{const}$.
	The relationship between $u_+$ and $v_-$ is given by the mirror trajectory as $u_+ = u_m(v_-)$ or, equivalently, $v_- = v_m(u_+)$.
	For an asymptotically lightlike mirror rays that start with $v > v_\mathcal{H}$ never get reflected; in a sense, they are lost to $\mathcal{I}^L_+$.
    }
	}
\label{fig:mirrors}
\end{figure*}

The moving mirror model is a 1+1d conformal field theory with a boundary (BCFT); a boundary condition is imposed along a timelike or a lightlike curve 
\hlch{(for an introduction see e.g. \cite{good2013time} and the textbooks \cite{Birrell:1982ix,Parker:2009uva}).}
In this paper we focus on the simplest case of the free boson CFT with a scalar field $\Phi$ and a boundary condition $\Phi = 0$ at the location of the mirror \cite{DeWitt:1975ys,Davies:1976hi,Davies:1977yv}.

The mirror location is taken to be preset externally.
Its trajectory can be specified in a variety of ways. In terms of the usual Minkowski coordinates $(t,z)$ we can specify the mirror position with the help of a function $t_m(z)$ or the inverse function $z_m(t)$ as
\begin{equation}
	t = t_m(z)  \quad
	\text{or} \quad
	z = z_m(t) \,.
\end{equation}
Equivalently, in terms of the light cone coordinates $(v,u)$ defined as
\begin{equation}
	v = t + z \,, \quad
	u = t - z \,,
\end{equation}
\hlch{
the mirror position can be set in terms of a function $u_m(v)$ or in terms of the inverse function $v_m(u)$ as
\begin{equation}
	u = u_m(v)  \quad
	\text{or} \quad
	v = v_m(u) \,.
\end{equation}
}
We will employ the usual convention that at late times the mirror goes to the left, i.e. \hlch{in the negative $z$ direction}.

There are two qualitatively different cases that depend on whether or not the mirror asymptotically approaches the speed of light.
If the mirror is always timelike, then all incident rays get reflected, see Fig.~\ref{fig:mirror_timelike}.
If the mirror is lightlike at late times, then it develops a horizon, as late rays never reach the mirror and do not get reflected, see Fig.~\ref{fig:mirror_lightlike}.
A comprehensive discussion of the moving mirror classification can be found in \cite{Akal:2022qei}.

Consider a ray that starts at the past lightlike infinity $\mathcal{I}^R_-$ with $v = v_-$ with a frequency $\omega_-$.
If this ray does get reflected from the mirror, then it travels to the future lightlike infinity $\mathcal{I}^R_+$ with $u = u_+$ (see Fig.~\ref{fig:mirrors}) and a red shifted frequency $\omega_+$.
The red shift is determined by a standard Doppler factor
\hlch{
\begin{equation}
	\frac{\omega_+}{\omega_-}
		= \frac{1 - |\dot{z}_m|}{1 + |\dot{z}_m|}
		= \dv{v_m}{u}
		= \left( \dv{u_m}{v} \right)^{-1} \,,
\label{mirror_redshift}
\end{equation}
}
where the dot denotes a derivative with respect to time $t$.

If the mirror is stationary or moves at a constant speed, we can always do a Lorenz boost to a frame where the mirror is at $z = z_m(t) \equiv 0$ at all times.
Such a boost does not change the scalar field energy-momentum tensor.
If we start in the vacuum state, we will always stay in the vacuum state.
We can reformulate this in the following way: if the red shift in Eq.~\eqref{mirror_redshift} is time-independent, there is no particle production.

However, if the mirror moves non-uniformly (i.e. the speed is not constant in time), the situation drastically changes.
In the CFT language, we can still perform a conformal map so that the mirror is at the origin, but the stress-energy tensor of the scalar field $\Phi$ become 
\hlch{non-trivial. 
In terms of the trajectory $v = v_m(u)$, one finds the energy flux at $\mathcal{I}^R_+$ \cite{Davies:1976hi}
\begin{equation}
	\left\langle T_{u u}\right\rangle
		=\frac{1}{24 \pi}\left[\frac{3}{2}\left(\frac{v_m^{\prime \prime}(u)}{v_m^{\prime}(u)}\right)^2-\frac{v_m^{\prime \prime \prime}(u)}{v_m^{\prime}(u)}\right] \,,
\label{mirror_T}
\end{equation}
where primes denote derivatives with respect to $u$.}

In the usual field theory language, while we start in the trivial vacuum $\ket{0}_\text{in}$, after the mirror's acceleration the state $\ket{0}_\text{in}$ is no longer a vacuum. Rather, it is a non-trivial linear combination of a new vacuum and excited states. This is reflected by non-trivial Bogolyubov coefficients.
The relevant equation is
(see e.g. \cite{Fabbri,Birrell:1982ix,good2013time,Good:2016atu})
\hlch{
\begin{equation}
	\beta^R_{\omega \omega^{\prime}}
		=\frac{1}{2 \pi} \sqrt{\frac{\omega^{\prime}}{\omega}} 
		\int\limits_{-\infty}^{v_\mathcal{H}} d v e^{-i \omega^{\prime} v -i \omega u_m(v)} \,.
\label{mirror_betas}
\end{equation}
}
Here, $v_\mathcal{H}$ is the position of the horizon ($v_\mathcal{H} = + \infty$ for a timelike mirror).
Using these Bogolyubov coefficients one can calculate the particle number registered at $\mathcal{I}^R_+$
\begin{equation}
	\expval{N_\omega} = \int\limits_0^\infty \diff \omega' \, |\beta^R_{\omega \omega^{\prime}}|^2
\end{equation}
as well as the corresponding energy
\begin{equation}
	\expval{E} = \int\limits_0^\infty \diff\omega \, \omega \expval{N_\omega}
	= \int\limits_0^\infty \diff\omega  \diff \omega' \, \omega |\beta^R_{\omega \omega^{\prime}}|^2 \,.
\end{equation}
\hlch{For further details we refer the interested reader to the textbooks \cite{Birrell:1982ix,Parker:2009uva}.}

\subsection{\hlch{Lightlike} shell in 3+1d}
\label{sec:setup_shell}

\begin{figure*}[t]
    \subfloat[Timelike shell that does not create a horizon.]{\label{fig:shell_timelike}\includegraphics[width=0.6\columnwidth]{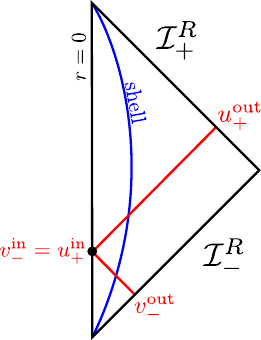}}
    \qquad
    \subfloat[\hlch{Lightlike shell} forms a horizon $\mathcal{H}$.]{\label{fig:shell_lightlike}\includegraphics[width=0.7\columnwidth]{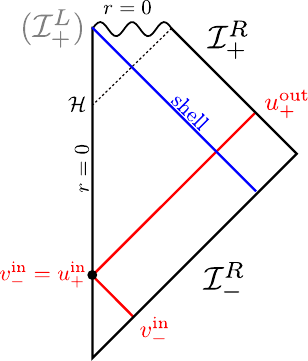}}
\caption{
	Penrose diagrams for 3+1 spacetime with simple shells.
	An incoming light ray starts at the past lightlike infinity $\mathcal{I}^R_-$ and propagates at $v = v_- = \text{const}$ (superscript \textquote{in} our \textquote{out} depending on the picture), then gets \textquote{reflected} at the origin $r=0$ and propagates to the future lightlike infinity $\mathcal{I}^R_+$ with $u^\text{out} = u_+^\text{out} = \text{const}$.
	The relationship between $u_+$ and $v_-$ is determined by the spacetime geometry.
	For a \hlch{lightlike shell} rays that start with $v > v_\mathcal{H}$ are lost to the singularity (analog of $\mathcal{I}^L_+$ in the mirror case).
	}
\label{fig:shells}
\end{figure*}

Our main model of gravitational collapse is going to be the dust shell \cite{Unruh:1976db,Wilczek:1993jn,Massar:1996tx}, see also \cite{Fabbri}.
We consider a shell of zero electrical charge and uniform surface energy density, with total energy $M$.
The shell is immersed in a 3+1 dimensional space with some underlying geometry (this background might be flat, or there might be e.g. a cosmological constant). 
Upon this, we consider a neutral massless scalar field $\Phi$ minimally coupled to gravity. 

The shell particles and the theory of gravity are taken as classical, while the field $\Phi$ will be considered as a quantum field propagating in the background of classical geometry induced by the shell; incorporation of the back reaction is left for a future work. 
The Hawking-like radiation that we want to study is the radiation of $\Phi$ quanta happening in the time-varying gravity background.

Generally, there are two qualitatively different possibilities.
In one situation there is no horizon; this happens when the shell is timelike and does not collapse, i.e. does not reach its Schwarzschild radius, see Fig.~\ref{fig:shell_timelike}.
In the other a horizon develops, as the shell collapses. This happens e.g. in the famous case of the \hlch{lightlike shell} and corresponding Vaidya spacetime, see Fig.~\ref{fig:shell_lightlike}.
%
In all of these cases one should differentiate between inside and outside of the shell, as the metric tensors are different in these regions.
Therefore, we denote the light cone coordinates inside the shell as $v^\text{in}$, $u^\text{in}$, and outside the shell $v^\text{out}$, $u^\text{out}$.
Precise relation of these coordinates to each other and to the usual spherical coordinates will be specified below.

In this paper we will focus on the case of null dust collapsing in an already non-trivial background geometry.
This case turns out to be much simpler than the timelike shell (the latter will be treated in a subsequent publication).
Indeed, consider a ray that starts at the past lightlike infinity $\mathcal{I}^R_-$ with $v^\text{out} = v^\text{out}_-$ with a frequency $\omega_-$ 
\hlch{(see Fig.~\ref{fig:shells}).}
If it does not hit the horizon, then it travels to the future lightlike infinity $\mathcal{I}^R_+$ with $u = u_+^\text{out}$ and a red shifted\footnote{Red shifted for a contracting shell. If the shell is expanding, then, generally speaking, there can be a blue shift.}
frequency $\omega_+$.
The simplicity of the \hlch{lightlike shell} case stems from the fact that such a ray crosses the shell exactly \textit{once}; in the timelike case there would be at least \textit{two} crossings, with the corresponding non-local effects severely complicating the analysis.

\vspace{10pt}

We define a ray tracing function $G(u)$ via
\begin{equation}
	v^\text{out}_-  = G( u_+^\text{out} ) \,.
\label{G_definition}
\end{equation}
%
As soon as the function $G(u)$ is known, the Doppler shift factor can be easily found
\begin{equation}
	\frac{\omega_+}{\omega_-}
		= \dv{G}{u} \,.
\label{GR_redshift}
\end{equation}
\hlch{
Comparing the Dopper shifts in the mirror case \eqref{mirror_redshift} and in the gravity case \eqref{GR_redshift}, one might wonder if the mirror trajectory $v_m(u)$ can be somehow identified with the gravitational ray tracing function $G(u)$. Below we will see that it is indeed true in a certain sense.
}

Hawking radiation in this setup is the production of the scalar field $\Phi$ quanta in such a time-varying background.
As was shown in \cite{Ford:1978ip}, 
the power radiated across a large sphere and registered at the future lightlike infinity $\mathcal{I}^R_+$ is given by
\hlch{
\begin{equation}
	P = \frac{1}{24 \pi}\left[\frac{3}{2}\left(\frac{G^{\prime \prime}(u)}{G^{\prime}(u)}\right)^2-\frac{G^{\prime \prime \prime}(u)}{G^{\prime}(u)}\right] \,.
\label{GR_power}
\end{equation}
}
In certain cases, the spectrum of Hawking radiation can also be computed; however, the calculations usually involve QFT on a curved spacetime background and often present challenges.

\section{Mirror-gravity correspondence in flat background}
\label{sec:wilczek}

In this section we are going to discuss the correspondence between the collapsing 3+1d geometry on one hand, and the moving mirror in 1+1d on the other hand.

We are going to start with a review of some known results on the mirror-gravity correspondence.
We will focus on the case when the black hole is modeled as a collapse of a thin shell made of null dust immersed in flat spacetime.
For such a shell one can construct a moving mirror that has (under approximations to be discussed below) the same radiation spectrum as that of the black hole under consideration.

The discussion presented here focuses on the key details that will be needed for further generalizations, which are also discussed below.

\begin{figure}[h]
    \centering
    \includegraphics[width=0.7\columnwidth]{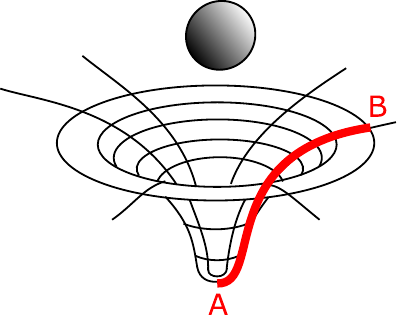}
\caption{
	Spacetime is curved in the presence of a massive object.
	The thick red line is a schematic representation of the distance to the origin.
	}
\label{fig:curved_manifold}
\end{figure}

\subsection{Lightlike shell collapse}

\hlch{Let us begin by briefly recounting the lightlike shell model.}
We take the bulk as the 3+1 Minkowski spacetime and introduce in it a collapsing \hlch{lightlike shell}.
The corresponding mirror was worked out in \cite{Wilczek:1993jn}.
In this setup, the space inside the shell is flat, while outside it is given by the Schwarzschild solution:
\begin{equation}
\begin{aligned}
	ds_\text{in}^2 &= dt_\text{in}^2 - dr^2 - r^2 d\Omega^2 \,, \quad r < r_\text{sh}(t) \,, \\
	ds_\text{out}^2 &= F(r) dt^2 - \frac{1}{F(r)} dr^2 - r^2 d\Omega^2  \,, \quad r > r_\text{sh}(t) \,.
\end{aligned}
\label{in_out_metric_wilczek}
\end{equation}
Here, $r_\text{sh}(t)$ is the radius of the collapsing shell, $M$ is the shell's energy, while the metric function $F$ is given by the standard expression
\begin{equation}
	F(r) = 1 - \frac{2M}{r} \,.
\end{equation}
Note that the time coordinate is different on the inside and on the outside; see Appendix~\ref{sec:inner_time} for the details.

We introduce the \hlch{Regge-Wheeler} tortoise coordinate $r_*(r)$ defined by
\begin{equation}
	\dv{r_*}{r} = \frac{1}{F(r)} \,.
\end{equation}
The light cone coordinates outside and inside the shell are then defined, \hlch{respectively}, as
\begin{equation}
	\begin{aligned}
		v^\text{out} = t + r_*(r) \,, \quad
			&u^\text{out} = t + r_*(r) \,; \\
		v^\text{in} = t_\text{in} + r \,, \quad
			&u^\text{in} = t_\text{in} + r \,.
	\end{aligned}
\label{lightcone_coord_in_out_def}
\end{equation}
Note that in terms of $r_*$, the $(t,r_*)$ components of the metric outside can be rewritten as $ds_\text{out}^2 = F(r) \left( dt^2 -  dr_*^2 \right)$.
%
\hlch{Note that this form of the metric is conformally flat. Therefore, it allows for solving the Klein-Gordon wave equation for a propagating massless scalar field.}

The correspondence with the moving mirror is made through the light cone coordinates.
In order to do that, we need to find the map that related the inner and outer regions of the shell.

First of all, one identifies $v^\text{in} = v^\text{out} \equiv v$.
This can be done because our shell is comprised of the incoming null dust.
\hlch{Indeed, following e.g. Sec.~29.2 of \cite{blau}, we write down the kinematic equation for such a shell as}
$v^\text{in} = v_\text{sh}^\text{in}$ or, equivalently, as $v^\text{out} = v_\text{sh}^\text{out}$ for some constants $v_\text{sh}^\text{in,out}$. By making appropriate shifts one can arrange for $v_\text{sh}^\text{in} = v_\text{sh}^\text{out}$, so that the identification $v^\text{in} = v^\text{out}$ is consistent on the collapsing shell.
There could still be a proportionality factor between them, but it turns out to be 1.

To determine the transformation function $u^\text{in} (u^\text{out})$, we study what has been called \textquote{the trajectory of the origin}.
From Eq.~\eqref{lightcone_coord_in_out_def} it follows that the point $r=0$ is determined by the equation $v^\text{in} = u^\text{in}$.
Requiring the shell's area to be a gauge invariant we can write
\begin{equation}
	r_* \left( \frac{v_\text{sh} - u^\text{in} (u^\text{out})}{2} \right) = \frac{v_\text{sh} - u^\text{out}}{2} \,,
\label{Wilczek_rstar_r}
\end{equation}
where $v_\text{sh} = t + r_\text{sh}(t)$ is the shell's location in spacetime.
Since for a given metric $F(r)$ the tortoise coordinate $r_*(r)$ is known,
Eq.~\eqref{Wilczek_rstar_r} determines the transformation function $u^\text{in} (u^\text{out})$.
%

\subsection{Statement of the correspondence}
\label{sec:statement_correspondence_flat}

\hlch{Now let us re-formulate the known mirror-gravity correspondence to be generalized below.
One} identifies this 3+1d collapsing geometry with a 1+1d moving mirror.
The mirror's trajectory is taken to be the trajectory of the origin; in a certain sense \cite{Wilczek:1993jn}, the distance from a one-dimensional observer to the mirror is taken to be the same as the distance to the origin as \textquote{seen} by the three-dimensional observer near a massive object, see Fig.~\ref{fig:curved_manifold} for a schematic representation.
In the spherical coordinates a ray sent to the origin is seen to be \textquote{reflected}, see Fig.~\ref{fig:shells}, which looks very similar to one side of the mirror Fig.~\ref{fig:mirrors}.

\hlch{To pass from the qualitative to the quantitative, we note that the ray tracing function $G(u^\text{out})$ (see Sec.~\ref{sec:setup_shell}) is the same as the mirror's trajectory function $v_m(u)$.}
They both map the incoming ray $u$-coordinate to the reflected outgoing ray $v$-coordinate.
The Doppler shift is given by the derivative of this function, see Eq.~\eqref{mirror_redshift} and Eq.~\eqref{GR_redshift};
the radiated energy flux is given by the Schwarzian derivative of this function, see Eq.~\eqref{mirror_T} and Eq.~\eqref{GR_power}.
The beta Bogolyubov coefficients, which are easily constructed for the mirror (see Eq.~\eqref{mirror_betas}), can also be constructed (to some extent) on the gravity side \cite{Hawking:1974sw,Fabbri}; the results are also in agreement with each other.

As a next step we note that the transformation function $u^\text{in} (u^\text{out})$ is actually the same as the ray tracing function $G$ defined in 
\hlch{Eq.~\eqref{G_definition}.}
This is evident from Fig.~\eqref{fig:shell_lightlike}: the ray starts inside the shell at $v^\text{in}_- \equiv v^\text{out}_-$, propagates at constant $v$, then reaches the origin where it starts propagating at constant $u^\text{in} = v^\text{in}_+$ (recall that $u^\text{in} = v^\text{in}$ for $r=0$) until it hits the shell. 
After the shell it propagates at constant $u^\text{out} = u^\text{out}_+$ such that $u^\text{in} = u^\text{in}(u^\text{out})$, see Eq.~\eqref{Wilczek_rstar_r}.
Tracing this ray back we obtain
\begin{equation}
	v^\text{out}_- = u^\text{in}(u^\text{out}_+) \,.
\label{v_u_lightlike}
\end{equation}
One can compare Eq.~\eqref{v_u_lightlike} with the definition of the $G$-function Eq.~\eqref{G_definition}.

All this allows to present the corresponding mirror's trajectory.
\hlch{
Given the transformation function $u^\text{in} (u^\text{out})$, one takes the mirror's trajectory in the form $v = v_m(u)$ (see Sec.~\ref{sec:setup_mirror}) as
\begin{equation}
	v_m(u) = u^\text{in}(u^\text{out} = u) \,.
\label{Wilczek_p}
\end{equation}
}
Identification \eqref{Wilczek_p} gives a mirror that produces radiation with the same spectrum as the corresponding black hole, assuming the geometric optics approximation on the gravity side (see e.g. \cite{Hawking:1974sw,Ford:1978ip,Fabbri}). Moreover, it captures only the $s$-wave radiation, hence the assumption of the spherical symmetry.

\subsection{Possible generalizations} 

Let us list a few key assumptions of the above analysis.
\begin{enumerate}
\item The spacetime under consideration is spherically symmetric. \label{item-sph}
\item The collapsing shell is lightlike, i.e. it collapses at the speed of light. \label{item-lightlike}
\item The spacetime inside the shell is flat. \label{item-inflat}
\item The spacetime outside the shell is given by the Schwarzschild solution. \label{item-schw}
\end{enumerate}
One can immediately generalize assumption \ref{item-schw} to a charged black hole \cite{good2020particle}, but beyond that one has to proceed with care.
For example, the cases of Kerr \cite{Good:2020fjz} and Schwarzschild -- de Sitter \cite{Fernandez-Silvestre:2021ghq} spacetimes require relaxing of the assumptions \ref{item-sph} and \ref{item-inflat}-\ref{item-schw}, \hlch{respectively}.

Below we are going to completely relax assumptions \ref{item-inflat} and \ref{item-schw} and consider the shell collapsing in an already non-trivial background. 
We will also comment a possible way to relax assumption \ref{item-sph} and consider an example of a spacetime that is only axially symmetric.
Generalization to timelike shells in connection to the mirror is going to be presented in a subsequent paper.

\section{The mirror and the non-trivial background geometry}
\label{sec:curvedback}

\hlch{In this Section we are going to implement the generalizations announced above and derive the main result of this paper.}
First of all, when we introduce a dust shell in an already non-trivial spacetime (for example, with a cosmological constant present), the metric inside the shell is no longer flat.
Instead of Eq.~\eqref{in_out_metric_wilczek} we will now have for the metric inside and outside the shell 
\begin{equation}
\begin{aligned}
	ds_\text{in}^2 &= F_\text{in}(r) dt_\text{in}^2 - \frac{1}{F_\text{in}(r)} dr^2 - r^2 d\Omega^2 \,, \quad r < r_\text{sh}(t)  \,; \\
	ds_\text{out}^2 &= F_\text{out}(r) dt^2 - \frac{1}{F_\text{out}(r)} dr^2 - r^2 d\Omega^2 \,, \quad  r > r_\text{sh}(t) \,.
\end{aligned}
\label{in_out_metric_lightlike_general}
\end{equation}
%
Note that the outside metric $F_\text{out}(r)$ is also no longer Schwarzschild but rather some fairly arbitrary function (of course, we assume it to be smooth and positive).

Since the inside is no longer flat, we need to introduce two corresponding tortoise coordinates instead of one:
\begin{equation}
	\dv{ r_{*\text{in}} }{r} = \frac{1}{ F_\text{in}(r) } \,, \quad
	\dv{ r_{*\text{out}} }{r} = \frac{1}{ F_\text{out}(r) } \,.
\label{in_out_metric_tortoise_general}
\end{equation}
%
The spacetime in coordinates $( t_\text{in}, r_{*\text{in}} )$ on the inside, and $( t, r_{*\text{out}} )$ on the outside is conformally flat, which motivates introduction of the lightlike coordinates
\begin{equation}
\begin{aligned}
	v^\text{in}  &= t_\text{in} + r_{*\text{in}} \,, \quad u^\text{in}  &= t_\text{in} - r_{*\text{in}} \,; \\
	v^\text{out}  &= t + r_{*\text{out}} \,, \quad u^\text{out}  &= t - r_{*\text{out}} \,.
\end{aligned}
\label{in_out_metric_lightcone}
\end{equation}
We still can identify $v^\text{in} = v^\text{out} \equiv v$
and describe the ingoing \hlch{lightlike shell} by $v = v_\text{sh} = \text{const}$.

The tortoise coordinates defined via Eq.~\eqref{in_out_metric_tortoise_general} are determined up to an integration constant.
While for the outer coordinate $r_{*\text{out}}$ this constant does not matter (it just effectively shifts $u^\text{out}$), the inner coordinate is fixed by requiring that $v^\text{in} = u^\text{in}$ at the origin $r=0$, which implies
\begin{equation}
	r_{*\text{in}} (r = 0) = 0 \,.
\label{in_tortoise_notmalization}
\end{equation}

Following the flat-background prescription, we still want to match the $r$ coordinate at the shell. From Eq.~\eqref{in_out_metric_lightcone} we infer that on the shell $r = r_\text{sh}$
\begin{equation}
	\begin{cases}
		r_{*\text{in}} &= \frac{ v_\text{sh} - u^\text{in} }{2} \,, \\
		r_{*\text{out}} &= \frac{ v_\text{sh} - u^\text{out} }{2} \,,
	\end{cases}
	\quad
	\text{on the shell}
\end{equation}
from which follows a generalization of Eq.~\eqref{Wilczek_rstar_r}
\begin{equation}
	r_{*\text{out}} \left[
		r_{*\text{in}}^{-1} \left(
			\frac{v_\text{sh} - u^\text{in} (u^\text{out})}{2} 
		\right)
	\right] = \frac{v_\text{sh} - u^\text{out}}{2} \,.
\label{in_out_metric_rstar_r}
\end{equation}
Note the appearance of the inverse function $r_{*\text{in}}^{-1}$.
\hlch{
The argument of Sec.~\ref{sec:statement_correspondence_flat} can be repeated, and we identify the mirror's trajectory $v_m(u) = u^\text{in} (u)$. 
}

\hlch{
Therefore, the generalized prescription can be formulated as
\begin{equation}
	r_{*\text{out}} \left[
		r_{*\text{in}}^{-1} \left(
			\frac{v_\text{sh} - v_m(u)}{2} 
		\right)
	\right] = \frac{v_\text{sh} - u}{2} \,.
\label{in_out_metric_recipe_pm}
\end{equation}
An alternative prescription for the same trajectory but in the form $u = u_m(v)$ would
\begin{equation}
	r_{*\text{out}} \left[
		r_{*\text{in}}^{-1} \left(
			\frac{v_\text{sh} - v}{2} 
		\right)
	\right] = \frac{v_\text{sh} - u_m(v)}{2} \,.
\label{in_out_metric_recipe_fm}
\end{equation}
}
The mirror's horizon is then located at $v_\mathcal{H} = v_\text{sh} - 2 r_{*\text{in}} (r_\mathcal{H})$, where $r = r_\mathcal{H}$ is the position of the collapsed BH horizon in this geometry (in the simple Schwarzschild case $r_\mathcal{H} = 2 M$).
Note that if the inside is actually flat ($r_{*\text{in}}(r) = r$) then Eq.~\eqref{in_out_metric_recipe_pm} reduces to the known prescription Eq.~\eqref{Wilczek_rstar_r}.

What we have done here is nothing but the generalization for the ray-tracing function $G(u)$.
The fundamental principle --- identification of the ray-tracing functions on the mirror side and on the gravity side --- remains unchanged.
\hlch{
Formula \eqref{in_out_metric_recipe_pm} gives the trajectory of the mirror corresponding to a given collapsing geometry;
this means that the Fulling-Davies-Unruh radiation coming from this mirror is the same as the Hawking radiation from the given collapsing geometry (in the geometric optics approximation).
}

\section{Applications}
\label{sec:examples}

Now let us apply the prescription derived in the previous section to some interesting cases.

\subsection{Schwarzschild -- de Sitter}

The case of a black hole in de Sitter universe with a cosmological constant $\Lambda > 0$ is interesting because of several reasons.
Our Universe is also expanding, and it was particularly important during the inflation \cite{Bousso:1997wi,Chao:1997em}.
Moreover, the cosmological horizon by itself also produces radiation with thermal characteristics \cite{Gibbons:1977mu}.
Such spacetimes are also interesting from the point of view of holography, see e.g. \cite{Marolf:2010tg} and references therein.

From the QFT perspective, thermodynamics of the Schwarzschild -- de Sitter (SdS) spacetime was first studied in \cite{Gibbons:1977mu}, followed by many others (see e.g. \cite{Kastor:1993mj,Bhattacharya:2018ltm} and references therein), and not only in 3+1 dimensions \cite{Kanti:2014dxa,Kanti:2017ubd,Pappas:2016ovo}.
The analysis of quantum fields in curved spacetimes such as SdS presents difficulties which are sometimes hard to overcome.
From this perspective the duality between spacetime geometries and moving mirrors seems attractive, as it allows one to employ QFT in flat 1+1d spacetime with a boundary to study properties (such us the radiation and its thermality) of higher-dimensional theories with gravity.

In \cite{Fernandez-Silvestre:2021ghq,Fernandez-Silvestre:2022gqn} the authors studied a mirror that corresponds to the following geometry: collapsing shell of null dust, positive cosmological constant $\Lambda > 0$ outside of the shell, and $\Lambda = 0$ inside the shell
\hlch{
so that the metric inside the shell is flat.
As we can see, in such a setup $\Lambda$ is actually not a constant, but varies over the spacetime.
Here we consider a more realistic scenario with the cosmological constant is positive and actually remains a constant
} everywhere, both inside and outside the shell.

\subsubsection{Mirror's trajectory}

The metric of the 3+1-dimensional SdS spacetime is of the form Eq.~\eqref{in_out_metric_lightlike_general} with the metric functions
\begin{equation}
	F_\text{in}(r) = 1 - \frac{r^2}{L^2} \,, \quad
	F_\text{out}(r) = 1 - \frac{r^2}{L^2} - \frac{2M}{ r } \,.
\label{F_in_out_SdS}
\end{equation}
The cosmological constant is related to the parameter $L$ as $L=\sqrt{3/\Lambda}$.
%
%
%
%
%
From Eq.~\eqref{in_out_metric_tortoise_general} and Eq.~\eqref{F_in_out_SdS} we find the inner tortoise coordinate and it's inverse:
\begin{equation}
\begin{aligned}
	r_{*\text{in}}(r) &= \frac{1}{2} L \ln( \frac{L + r}{L - r} ) = L \tanh[-1](\frac{r}{L}) \,, \\
	r_{*\text{in}}^{-1}(R) &= L \tanh( \frac{R}{L} ) \,.
\end{aligned}
\label{SdS_tortoise_inner}
\end{equation}
Note the normalization $r_{*\text{in}}(0) = 0$, which ensures $v^\text{in} = u^\text{in}$  at the origin $r=0$, see Sec.~\ref{sec:curvedback}.
The outer tortoise coordinate reads
\begin{equation}
\begin{aligned}
	r_{*\text{out}}(r)
		&= \frac{1}{2\kappa_B}\ln\left(\frac{r}{r_B} - 1\right) \\
		&-\frac{1}{2\kappa_C}\ln\left(1 - \frac{r}{r_C}\right) \\
	    &+ \left( \frac{1}{2\kappa_C} - \frac{1}{2\kappa_B} \right) \ln\left( 1 + \frac{r}{ r_B + r_C } \right) \,.
\end{aligned}
\label{SdS_tortoise_outer}
\end{equation}
%
In Eq.~\eqref{SdS_tortoise_outer}, the $r_B$, $r_C$ are the two Killing horizons related to the black hole and de Sitter cosmology respectively,
\begin{equation}
\begin{aligned}
	r_B &= \frac{2L}{\sqrt{3}}\cos\left[\frac{1}{3}\cos^{-1}\left(\frac{3\sqrt{3}M}{L}\right)+\frac{\pi}{3}\right] \,, \\
	r_C &= \frac{2L}{\sqrt{3}}\cos\left[\frac{1}{3}\cos^{-1}\left(\frac{3\sqrt{3}M}{L}\right)-\frac{\pi}{3}\right] \,.
\end{aligned}
\label{SdS_horizons}
\end{equation}
The corresponding surface gravities are given by
\hlch{
\begin{equation}
\begin{aligned}
	\kappa_B&= \frac{(r_C-r_B)(2r_B+r_C)}{2L^2r_B} \,,\\
	\kappa_C&= \frac{(r_C-r_B)(2r_C+r_B)}{2L^2r_C} \,.
\end{aligned}
\end{equation}
%
}

Now, applying the recipe Eq.~\eqref{in_out_metric_recipe_fm} (and setting $v_\text{sh}=0$ for convenience) we obtain the SdS trajectory:
\hlch{
\begin{equation}
\begin{aligned}
	u_m& (v) = - \frac{1}{\kappa_B}\ln\left[ - \frac{L}{r_B} \tanh( \frac{v}{2L} ) - 1\right] \\
		&+ \frac{1}{\kappa_C}\ln\left[1 + \frac{L}{r_C} \tanh( \frac{v}{2L} ) \right] \\
	    &+ \left( \frac{1}{\kappa_B} - \frac{1}{\kappa_C} \right) \cdot \ln\left[ 1 - \frac{L}{ r_B + r_C } \tanh( \frac{v}{2L} ) \right] \,.
\end{aligned}
\label{SdS_mirror}
\end{equation}
}
Formula \eqref{SdS_mirror} gives the mirror corresponding to the SdS black hole.

\subsubsection{Analysis of the dynamics}

Our trajectory in Eq.~\eqref{SdS_mirror} can be compared to the trajectory derived in \cite{Fernandez-Silvestre:2021ghq} (see Eq.~(20-22) in that paper).
There are several distinct features.
We start by  checking the limiting cases.

For a vanishing cosmological constant $\Lambda \to 0$, i.e. $L \to \infty$, we have $r_B \approx 2M$, $\kappa_B \approx 1/(4M)$ and $r_C \approx L$, $\kappa_C \approx 1/L$.
Decomposing the logs in Eq.~\eqref{SdS_mirror} we obtain
\hlch{
\begin{equation}
	u_m(v) \approx v - 4M \ln(- v / 2 M) + \text{const} \,,
\end{equation}
}
which is the Schwarzschild mirror and agrees with \cite{Fernandez-Silvestre:2021ghq,Wilczek:1993jn}.

However, the limit of empty dS space turns out to be different from \cite{Fernandez-Silvestre:2021ghq}.
In this limit we take $M \to 0$, then $r_B \approx 2M \to 0$, $\kappa_B \approx 1/(4M) \to \infty$ and $r_C \approx L$, $\kappa_C \approx 1/L$. 
The trajectory \eqref{SdS_mirror} reduces to
\hlch{
\begin{equation}
	u_m(v) = v + \text{const} \,,
\label{sds_zeroM_traj}
\end{equation}
}
which is nothing but the stationary mirror, and there is no radiation in this case.
The reasons for such behavior will be discussed below.

Now let us turn to the radiation spectrum.
In principle, the spectrum of the radiation (Davies-Fulling-Unruh effect) can be obtained by plugging the trajectory \eqref{SdS_mirror} into the integral \eqref{mirror_betas}.
This is already a strong result, as it presents the spectrum as a one-dimensional integral of an analytic function easily tractable by numerical analysis.

The most important contributions to the spectrum can be easily derived analytically in the near-horizon approximation.
The trajectory equation \eqref{SdS_mirror} diverges at the values of $v = v_{\mathcal{H},B}$ and $v = v_{\mathcal{H},C}$ determined by the equations 
\begin{equation}
	\tanh( \frac{ v_{\mathcal{H},i} }{2L} ) = - \frac{r_i}{L} \,, \quad 
	i = B,C \,.
\end{equation}
We assume $3 \sqrt{3} M / L < 1$, so that $r_B < r_C$, and we are below the Nariai limit.
In this case, simple analysis of Eq.~\eqref{SdS_horizons} shows that $0 < r_B < L/\sqrt{3}$ and $L/\sqrt{3} < r_C < L$, and $ v_{\mathcal{H},C} < v_{\mathcal{H},B} < 0 $.
(There is also a non-physical third \textquote{horizon}).

\hlch{
When $v$ approaches one of the horizon values, the mirror trajectory \eqref{SdS_mirror} simplifies to a single log\footnote{There are also unimportant constants which make the logs dimensionless.}:
\begin{equation}
\begin{aligned}
	v \to v_{\mathcal{H},B} &\Rightarrow u_m(v) \approx - \frac{1}{\kappa_B} \ln \big[ - (v - v_{\mathcal{H},B}) ] \,, \\
	v \to v_{\mathcal{H},C} &\Rightarrow u_m(v) \approx - \frac{1}{\kappa_C} \ln \big[ v - v_{\mathcal{H},C} ] \,.
\end{aligned}
\label{SdS_traj_nearhor}
\end{equation}
}
These are nothing but examples of the Carlitz-Willey mirror \cite{carlitz1987reflections}.
For this mirror the radiated particle spectrum is known exactly: it is the 1+1 Planck distribution with a well-defined temperature $T$,
\begin{equation}
\begin{aligned}
		N_{\omega \omega', i} &= |\beta_{\omega \omega'}|^2 = \frac{1}{ 2 \pi \kappa_i \omega' } \frac{1}{ e^{2 \pi \omega / \kappa_i} - 1 } \,, \\
		T_i &= \frac{\kappa_i}{2 \pi} \,, \quad
		i = B,C \,.
\end{aligned}
\end{equation}
%
Thus we obtain the thermal radiation with the temperature that agrees with known results.
The total radiation spectrum is a sum of the two found temperature curves plus some residual non-thermal radiation.
The latter piece can be prominent only during the intermediate times, but is negligible at the late times, when the black hole horizon forms.


\subsubsection{Discussion of the results}

\hlch{
In this example we saw that in order to derive the spectrum of the black hole radiation it is not necessary to employ the full machinery of QFT in curved spacetime.
Instead, one can build an analogous mirror model and arrive at this spectrum with less effort.
}

The main results, namely, the two-temperature spectrum and the expressions for the temperatures, agree with the literature.
However, there is one seeming peculiarity: the fact that in the limit of empty $dS$ spacetime, $M \to 0$, we get a mirror that does not radiate, \hlch{see \eqref{sds_zeroM_traj}}.
This happened because in our dynamical \hlch{lightlike shell} collapse actually models only the black hole radiation.
The asymptotically de Sitter space is static in this model, and it is not captured by the mirror, which is inherently dynamical.
The mirror corresponding to the eternally empty dS space is a static mirror and does not radiate.

\vspace{10pt}

\hlch{
One might wonder, why then we see radiation with the cosmological temperature $T_C$ at all.
It turns out that this radiation is also caused by the shell's movement, but in the distant past.
Let us explain this in more detail.
}

\hlch{
The lightlike shell in this construction, before collapsing to form a black hole, comes from the past lightlike infinity.
Since the cosmological horizon is static in this model, the lightlike shell actually started from that horizon, meaning that in the limit $t \to - \infty$ the shell's radius actually approaches the de Sitter horizon $r_C$, see Eq.~\eqref{SdS_horizons}.
There is then thermal radiation dynamically generated by this shell in the past.
}

\hlch{
In other words, the timelike shell approaches the black hole horizon in the future, and the cosmological horizon in the past.
The corresponding moving mirror also displays the two horizons, the future and the past one, see Eq.~\eqref{SdS_traj_nearhor}.
And in the limit $M \to 0$ there is no shell and, therefore, no dynamically generated radiation.
}

\vspace{10pt}

\hlch{The fact that the moving mirror models do not capture static spacetimes is also true for the pure Schwarzschild black hole case.}
All known constructions of this sort do not deal with eternal black holes; instead, they involve some dynamical models for the black hole collapse, for example, the shell of null dust.
This inherent property of the mirror-gravity correspondence is not a serious drawback, in our view, as realistic astrophysical black holes are not eternal, but rather form in a collapse, either in the process of the stellar evolution or in fluctuations shortly after the Big Bang.


One can also consider a model that dynamically incorporates the dS \hlch{cosmological horizon as well.
One such} construction was proposed in \cite{Good:2020byh}, where a bubble of de Sitter vacuum was expanding into the flat Minkowski exterior.
Incorporation of another collapsing \hlch{lightlike shell} into this spacetime is left for a future work.

\subsection{BTZ}
\label{sec:BTZ}

The Ba\~nados--Teitelboim--Zanelli (BTZ) black hole \cite{Banados:1992wn} is an important solution of the (topological) gravity in 2+1-dimensional AdS spacetime.
Since it is possible to model this black hole as a collapse of a null dust shell, the mirror dual can be easily constructed.

This example also uncovers some new advantages of the moving mirror approach.
Quantum theory in the BTZ spacetime has several distinct features.
\begin{enumerate}
\item 
AdS space, if taken as-is, admits closed timelike curves. To rectify the situation one should pass to the universal covering space (CAdS).

\item 
AdS space has a timelike boundary, though which information can escape. In other words, if a light ray is emitted from the origin, it reaches the infinity in a finite coordinate time.

\item 
BTZ black hole is topological in the sense that it is not a compact object curving the spacetime.\footnote{This feature actually does not present problems in QFT, but we list it for completeness.}
Rather, BTZ is just a discrete quotient of the empty AdS$_3$. 
It was first noted in 2+1 dimensions \cite{Banados:1992wn,Banados:1992gq}, then generalized to 3+1 \cite{Aminneborg:1996iz} and to higher-dimensional spaces \cite{Banados:1997df}, see also \cite{Steif:1995zm} for a detailed explanation.

\end{enumerate}
Below we will see that all these features are easily addressed in the moving mirror approach.

\begin{figure}[h]
    \centering
    \includegraphics[width=0.2\textwidth]{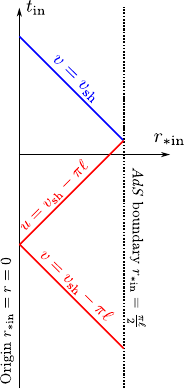}
    \caption{
    	\hlch{Lightlike} shell in AdS background.
    	Coordinates $t_\text{in}$, $r_{*\text{in}}$ are used.
    	Due to the presence of $AdS$ boundary, rays that are emitted too early at $v < v_\text{sh} - \pi\ell$ never encounter the shell.
    }
    \label{fig:btz_ray}
\end{figure}

\subsubsection{Mirror's trajectory}

In our setup we consider a \hlch{lightlike shell} of effective mass $M+1$ collapsing on a pre-existing AdS$_3$ space\footnote{The AdS$_3$ space can be viewed as a BTZ black hole with mass $M_0=-1$ \cite{Banados:1992wn}. We were not able to construct a mirror corresponding to a shell that collapses on the $M_0=0$ configuration, as in this case the inner tortoise coordinate would diverge at the origin, and the normalization discussed in Sec.~\ref{sec:curvedback} cannot be performed.}. 
Similar setup --- AdS inside, BTZ outside --- has been also considered in the gravity side \cite{Hyun:1994na}; the difference is that the authors of the latter paper consider a quantum scalar field with the conformal coupling to gravity, while here we consider the minimal coupling.
This however does not affect the radiation temperature.

The BTZ black hole can be modeled by a \hlch{lightlike shell} collapse \cite{Husain:1994xa,Virbhadra:1994xz,Chan:1994rs}.
In the Eddington-Finkelstein coordinates the Vaidya-like metric in this case is
\begin{equation}
	ds^2 = - \left( m(v) + \frac{r^2}{\ell^2} \right) dv^2 - 2 dv dr - r^2 d\phi^2 \,,
\end{equation}
where the mass function is given by
\begin{equation}
	m(v) = \begin{cases}
		-1, v < v_\text{sh} \\
		M, v > v_\text{sh} \,.
	\end{cases}
\end{equation}
\hlch{ Note that the negative mass value $m=-1$ here is in fact physical. As was noted already in the original article \cite{Banados:1992wn}, a generic negative value of $m$ leads to naked singularities, but the value $m=-1$ is exceptional and gives the empty AdS spacetime.
}

In spherical coordinates, the metric inside and outside of the shell thus is the 2+1D analog of \eqref{in_out_metric_lightlike_general},
\begin{equation}
\begin{aligned}
	ds_\text{in}^2 &= F_\text{in}(r) dt_\text{in}^2 - \frac{1}{F_\text{in}(r)} dr^2 + r^2 d\phi^2  \,, \\
	ds_\text{out}^2 &= F_\text{out}(r) dt^2 - \frac{1}{F_\text{out}(r)} dr^2 + r^2 d\phi^2  \,.
\end{aligned}
\label{in_out_metric_lightlike_btz}
\end{equation}
with the metric functions
\begin{equation}
	F_\text{in}(r) = 1 + \frac{r^2}{\ell^2} \,, \quad
	F_\text{out}(r) = - M + \frac{r^2}{\ell^2} \,.
\end{equation}
The inner tortoise coordinate is then given by
\begin{equation}
\begin{aligned}
	r_{*\text{in}}(r) &=  \ell \tan[-1](\frac{r}{\ell})  \,, \\
	r_{*\text{in}}^{-1}(R) &= \ell \tan( \frac{R}{\ell} ) \,.
\end{aligned}
\label{BTZ_tortoise_inner}
\end{equation}
The outer tortoise coordinate is also simply found
\begin{equation}
\begin{aligned}
	r_{*\text{out}}(r) &= - \frac{\ell}{\sqrt{M}} \tanh[-1](\frac{\ell \sqrt{M} }{r})  \,,\\
	r_{*\text{out}}^{-1}(R) &= \frac{ \ell \sqrt{M} }{ \tanh( \frac{- R\sqrt{M}}{\ell} ) } \,.
\end{aligned}
\label{BTZ_tortoise_outer}
\end{equation}

Plugging Eq.~\eqref{BTZ_tortoise_inner} and Eq.~\eqref{BTZ_tortoise_outer} into the mirror-gravity recipe Eq.~\eqref{in_out_metric_recipe_pm} or Eq.~\eqref{in_out_metric_recipe_fm}, after some algebra we get the mirror trajectory:
%
\hlch{
\begin{equation}
	u_m(v) = v_\text{sh} + \frac{2\ell}{\sqrt{M}} \tanh[-1]( \frac{\sqrt{M}}{ \tan( \frac{v_\text{sh} - v}{2\ell} ) } ) \,,
\label{BTZ_f}
\end{equation}
}
%
Here, $v_\text{sh}$ is the constant specifying the shell's position in the $(u,v)$ frame.

\subsubsection{AdS peculiarities in the mirror}

Now, let us see how the difficulties outlined at the beginning of this section are resolved in the moving mirror formalism.
\begin{enumerate}
\item 
Presence of closed timelike curved in AdS manifested itself in the fact that the expression \eqref{BTZ_f} is a periodic function in $v$.
To rectify this, it suffices to simply consider any one particular branch of this function. 

\item
AdS boundary manifested itself in the fact that, in addition to the expected black hole future horizon, the expression \eqref{BTZ_f} has a past horizon.
It turns out that, in order to rectify this, we should take into account only a half of the trajectory.
Let us consider this in detail.
\end{enumerate}

From the AdS tortoise coordinate in Eq.~\eqref{BTZ_tortoise_inner} one can see that it takes $\pi\ell$ amount of time for a light ray to cross the whole of AdS.
Therefore, a light ray emitted too early, i.e. at $v < v_\text{crit} = v_\text{sh} - \pi\ell$, never actually encounters the \hlch{lightlike shell} propagating at $v = v_\text{sh}$, see Fig.~\ref{fig:btz_ray}.
Therefore such rays should not be taken into account, and in the moving mirror one should only consider $v > v_\text{crit}$.

The future and the past horizons are defined by the equations 
\hlch{
\begin{equation}
	u_m ( v_{\mathcal{H},\text{past}} ) = - \infty \,, \quad
	u_m ( v_{\mathcal{H},\text{fut}} ) = + \infty 
\end{equation}
}
that give (choosing a particular branch)
\begin{equation}
\begin{aligned}
	v_{\mathcal{H},\text{past}} &= v_\text{sh} + 2 \ell \tan[-1]( \sqrt{M} ) - 2 \pi \ell \,, \\
	v_{\mathcal{H},\text{fut}} &= v_\text{sh} - 2 \ell \tan[-1]( \sqrt{M} ) \,.  \\
\end{aligned}
\label{btz_horizon_v}
\end{equation}
As it turns out, using only $v > v_\text{crit} = v_\text{sh} - \pi\ell$ cuts out exactly a half of this trajectory.
The other half at $v < v_\text{crit}$ is just a fictitious reflection of it.

Plugging the resulting trajectory in the beta Bogolyubov integral with the appropriate limits, we finally arrive at the formula for the BTZ radiation spectrum,
\hlch{
\begin{equation}
	\beta^R_{ \omega \omega' }
		=\frac{1}{2 \pi} \sqrt{\frac{ \omega' }{ \omega }} 
		\int\limits_{v_\text{sh} - \pi\ell}^{ v_{\mathcal{H},\text{fut}}  } d v \,
			e^{-i \omega' v -i \omega u_m(v)} \,.
\label{btz_betas}
\end{equation}
}

\subsubsection{Radiation and temperature}

Now let us actually analyze the BTZ mirror trajectory and investigate the radiation spectrum.

First of all we note that, just as it was in the previous example, in the limit of empty AdS space (i.e. with a shell of vanishing mass $(M+1) \to 0$ we obtain a stationary mirror:
\hlch{
\begin{equation}
	u_m(v) = - 2\ell \tan^{-1} \left[ \cot( \frac{v}{2\ell} )  \right]
		= v - \pi \ell \,.
\label{BTZ_f_M=-1}
\end{equation}
}
There is no radiation in this case.
This is emphasizes the fact that our current model describes dynamically only the black hole. 
The AdS bulk is static, and possible radiation from the AdS boundary is not captured in this model.

Next, we want to extract the temperature of the thermal radiation spectrum.
Again, in principle the spectrum is given by the exact formula \eqref{btz_betas}.
The temperature can be easily extracted in the near-horizon approximation, when the trajectory Eq.~\eqref{BTZ_f} simplifies to
\hlch{
\begin{equation}
	u_m(v) \approx - \frac{\ell}{ \sqrt{M} } \ln \big[- (v - v_{\mathcal{H},\text{fut}} ) \big] + \text{const} \,.
\label{BTZ_f_ellhalf_nearhor}
\end{equation}
}
This gives the particle spectrum
\begin{equation}
	N_{\omega \omega'} =
	|\beta^R_{ \omega \omega' }|^2 \approx \frac{1}{2 \pi \kappa \omega' } \frac{1}{ e^{2 \pi \omega / \kappa} - 1 } \,, \quad
	\kappa = \frac{\sqrt{M} }{ \ell } \,.
 \label{BTZ-beta-identified}
\end{equation}
Thus the spectrum is Planck with the temperature
\begin{equation}
	T_\text{BTZ} = \frac{\sqrt{M} }{ 2 \pi \ell } \,.
\end{equation}
The spectrum and the temperature just obtained coincide with known results from the gravity side \cite{Banados:1992wn,Hyun:1994na}.


\subsubsection{Note on a spinning black hole}
\label{sec:spin}

To conclude this section, we want to make a remark concerning the case of BTZ with angular momentum.
Although the prescription described in this paper is not a priori guaranteed to work in this case, it nevertheless gives the right temperature.

For a BTZ black hole with mass $M$ and angular momentum $J$, the metric has the form \cite{Banados:1992wn}
\begin{equation}
	ds_\text{out}^2 = F_\text{out}(r) dt^2 - \frac{1}{F_\text{out}(r)} dr^2 + r^2 \left(  d\phi - \frac{J}{2 r^2} dt \right)^2
\label{BTZ_J_metric}
\end{equation}
with
\begin{equation}
	F_\text{out}(r) = - M + \frac{r^2}{\ell^2} + \frac{J^2}{4 r^2} ,.
\end{equation}
Introducing
\begin{equation}
	r_\pm = \ell \sqrt{ \frac{M}{2} \left( 1 \pm \sqrt{1 - \frac{J}{M \ell} } \right) } \,,
\end{equation}
one can write the expression for the tortoise coordinate in a convenient form
\begin{equation}
	r_{*\text{out}}(r) = \frac{ \ell^2 }{ r_+^2 - r_-^2 } \left[ r_- \tanh[-1](\frac{r_-}{r}) - r_+ \tanh[-1](\frac{r_+}{r}) \right] \,.
\label{BTZ_J_tortoise_outer}
\end{equation}

Now let us take a \hlch{lightlike shell} collapsing to form this spinning BTZ black hole.
Assuming AdS metric inside, one can easily write down the mirror trajectory:
\hlch{
\begin{equation}
\begin{aligned}
	u_m(v) = v_\text{sh} 
		&+ \frac{ \ell^2 }{ r_+^2 - r_-^2 } \Bigg[ 
			r_- \tanh[-1](\frac{r_-}{ \tan( \frac{v_\text{sh} - v }{2\ell} ) }) \\
			&- r_+ \tanh[-1](\frac{r_+}{ \tan( \frac{ v_\text{sh} - v }{2\ell} ) }) 
		\Bigg] \,.
\end{aligned}
\label{BTZ_J_f}
\end{equation}
}
On the gravity side, the surface $r = r_+$ is the black hole horizon.
The corresponding mirror horizon $v_{\mathcal{H},\text{fut}}$ is determined by the equation $\tan( v / 2\ell ) = r_+$; in the near-horizon approximation the mirror becomes
\hlch{
\begin{equation}
	u_m(v) \approx - \frac{ \ell^2 r_+ }{ r_+^2 - r_-^2 } \ln(- (v - v_{\mathcal{H},\text{fut}} ) ) + \text{const} \,.
\end{equation}
}
This is again a Carlitz-Willey mirror, which gives thermal radiation spectrum with the temperature
\begin{equation}
	T_{\text{BTZ},J} = \frac{ r_+^2 - r_-^2 }{ 2 \pi \ell^2 r_+ } \,.
\label{BTZ_J_temperature}
\end{equation}
This again coincides with known results \cite{Banados:1992wn,Hyun:1994na}.

When the angular momentum is zero, $J=0$, the two radii become $r_+ = \ell \sqrt{M}$, $r_- = 0$, and we recover the results from the above.

In the extremal limit $J = M \ell$ we have $r_+ = r_- = \ell \sqrt{M/2}$, and the temperature in Eq.~\eqref{BTZ_J_temperature} drops to zero.
This does not mean that such a black hole does not radiate, it only means that the radiation spectrum is no longer thermal.
Indeed, the mirror trajectory in this case becomes
%
\hlch{
\begin{equation}
\begin{aligned}
	u_m(v) = v_\text{sh} 
		&- \frac{ \ell }{ \sqrt{ 2 M } } \Bigg[ 
			\tanh[-1]( x ) - \frac{1}{ x - 1/x }
		\Bigg] \,, \\
	x &\equiv \frac{ \ell \sqrt{M/2} }{ \tan( \frac{v_\text{sh} - v}{2\ell} ) } \,.
\end{aligned}
\label{BTZ_extr_f}
\end{equation}
}
In the near-horizon limit $x \to (1 \, - \! 0)$ the first term in the square brackets gives the expected log, but the second term has a pole and spoils thermality.
This is consistent with known results on the extremal Kerr black hole \cite{Rothman:2000mm}.

\vspace{10pt}

Now, the important question is, why has our prescription given the correct results, even though the spacetime in Eq.~\eqref{BTZ_J_metric} was not spherically symmetric?
The answer is that, although there is no strict spherical symmetry, the current setup is still spherically symmetric in a sort of a \textit{weak} sense, meaning that the tortoise coordinate in Eq.~\eqref{BTZ_J_tortoise_outer} still does not depend on the angle variable.

In the spinless case $J=0$ the meaning of the coordinate $r_{*\text{out}}(r)$ is clear: it describes the optical path of massless field waves.
When $J \neq 0$, this is not so straightforward, because massless particles' geodesics generically do not satisfy $\phi = \text{const}$.
However, consider a ray sent in the direction of the origin (the $s$-wave approximation).
From the BTZ geodesics equations \cite{Cruz:1994ir} at zero angular momentum ($L=0$ in the notation of that paper) it follows that
\begin{equation}
    \dv{\phi}{t} = \frac{J}{2 r^2} \,.
\end{equation}
Thus the last term of Eq.~\eqref{BTZ_J_metric} vanishes, and we are left with the spherically symmetric interval for the massless $s$-wave.
This property can be called $s$-wave angular (or spherical) symmetry.

Therefore we see that in the case of rotating BTZ black hole we still recover spherical symmetry in the $s$-wave sector, and that is the reason why the moving mirror picture still works.
\hlch{It would be interesting to rigorously analyze the lightlike geodesics and the mirror approach for the Kerr black hole in 3+1 dimensional spacetime; we leave this problem for another paper.}


\section{Conclusions}
\label{sec:conclusions}

In this paper we presented a generalization of the correspondence between black holes and moving mirrors.
\hlch{
This construction allows one to study certain aspects of black holes, like the radiation and temperature, by modeling such a black hole by a collapse of a shell made of null dust, and then constructing a  dynamical moving mirror in 1+1d that corresponds to the spacetime of interest.
This paper focuses on constructing a mirror that corresponds to a given black hole geometry. Once a mirror trajectory is constructed, the usual methods can be employed to study the beta Bogolyubov coefficients.}

\hlch{
The moving mirror approach is simpler as compared to the direct analysis of QFT in the curved spacetime of interest.
We have demonstrated this by studying the spectrum of the Hawking radiation of the Schwarzschild -- de Sitter and BTZ black holes by relating it to the Fulling-Davies-Unruh effect in 1+1d.
}

\hlch{
In this paper we focused on generalizing and testing the moving mirror approach.
}
Previously this correspondence was well-motivated only in spherically symmetric asymptotically flat cases.
\hlch{
Here we extended it 1) to the non-trivial (asymptotically non-flat) background geometry, and 2) we were also able to somewhat relax the spherical symmetry assumption: the spacetime does not have to be spherically symmetric, 
only the $s$-wave sector of massless particles' geodesics should have spherical symmetry.
It is plausible that this last assumption can also be lifted, allowing for a rigorous study of e.g. Kerr black holes via the moving mirror;
however, further investigation is needed to clarify this.
}

\hlch{
The next step in investigating the mirror-gravity duality is to find a field-theoretical proof of the correspondence between the beta Bogolyubov coefficients on both sides.
}

This correspondence is promising from the point of view of the information loss and other entropy-related questions.
It is also possible to include the back reaction into this \hlch{setup \cite{Oku:1979ey,Hotta:1994ha,Good:2022wpw,Xie:2023wvu,Fabbri}.}
We are going to pursue these avenues in the immediate future.
Another direction that will be pursued in the future works is incorporation of timelike shells and balls.
There are indications that such timelike objects might describe what has been called remnants.


\begin{acknowledgments} 

The author thanks Michael R.R. Good for very useful discussions and a reading of the manuscript draft.
This work is supported by the William I. Fine Theoretical Physics Institute at the University of Minnesota,
and also by the FY2021-SGP-1-STMM Faculty Development Competitive Research Grant No.\ 021220FD3951 at Nazarbayev University.

\end{acknowledgments}


\appendix

\section{The time coordinate inside the shell}
\label{sec:inner_time}

In this Appendix we make a brief note on the time coordinate in the flat inner region, $t_\text{in}$, see Eq.~\eqref{in_out_metric_wilczek}.
As the shell is lightlike, we have $ds^2 = 0$ on the shell.
From the metric Eq.~\eqref{in_out_metric_wilczek} it follows then
\begin{equation}
\begin{aligned}
	0 &= \left( \dv{ t_\text{in} }{t} \right)^2 - \left( \dv{r_\text{sh}(t)}{t} \right)^2  \\
	0 &= F(r_\text{sh}(t)) - \frac{1}{F(r_\text{sh}(t))} \left( \dv{r_\text{sh}(t)}{t} \right)^2 
\end{aligned}
\label{in_out_metric_wilczek_junction}
\end{equation}
from which we immediately obtain
\begin{equation}
	\dv{ t_\text{in} }{t} = F(r_\text{sh}(t))
\end{equation}
Therefore, the metric Eq.~\eqref{in_out_metric_wilczek} can be represented in the same coordinates inside and outside of the shell as
\begin{equation}
\begin{aligned}
	ds_\text{in}^2 &= h^2(t) dt^2 - dr^2 - r^2 d\Omega^2 \,, \quad r < r_\text{sh}(t) \,, \\
	ds_\text{out}^2 &= F(r) dt^2 - \frac{1}{F(r)} dr^2 - r^2 d\Omega^2 \,, \quad r > r_\text{sh}(t) \,, \\
        h^2(t) &= F^2(r_\text{sh}(t)) \,.
\end{aligned}
\label{in_out_metric_wilczek_camecoord}
\end{equation}
%
%
This demonstrates that, although the metric inside the shell is indeed flat, it is not the naive \textquote{$dt^2 - dr^2$}.

\bibliography{main}

\end{document}